\documentclass[twocolumn,aps,prb,superscriptaddress]{revtex4}
\pdfoutput=1
\usepackage{graphicx}
\usepackage{hyperref}
\begin{document}

\title{Coupling Superconducting Qubits via a Cavity Bus}

\author{J.~Majer}
\affiliation{Departments of Applied Physics and Physics, Yale University, New Haven, CT 06520}
\affiliation{These authors contributed equally to this work.}
\author{J.~M.~Chow}
\affiliation{Departments of Applied Physics and Physics, Yale University, New Haven, CT 06520}
\affiliation{These authors contributed equally to this work.}
\author{J.~M.~Gambetta}
\affiliation{Departments of Applied Physics and Physics, Yale University, New Haven, CT 06520}
\author{Jens ~Koch}
\affiliation{Departments of Applied Physics and Physics, Yale University, New Haven, CT 06520}
\author{B.~R.~Johnson}
\affiliation{Departments of Applied Physics and Physics, Yale University, New Haven, CT 06520}
\author{J.~A.~Schreier}
\affiliation{Departments of Applied Physics and Physics, Yale University, New Haven, CT 06520}
\author{L.~Frunzio}
\affiliation{Departments of Applied Physics and Physics, Yale University, New Haven, CT 06520}
\author{D.~I.~Schuster}
\affiliation{Departments of Applied Physics and Physics, Yale University, New Haven, CT 06520}
\author{A.~A.~Houck}
\affiliation{Departments of Applied Physics and Physics, Yale University, New Haven, CT 06520}
\author{A.~Wallraff}
\affiliation{Departments of Applied Physics and Physics, Yale University, New Haven, CT 06520}
\affiliation{Present address: Department of Physics, ETH Zurich, CH-8093 Z\"{u}rich, Switzerland}
\author{A.~Blais}
\affiliation{Departments of Applied Physics and Physics, Yale University, New Haven, CT 06520}
\affiliation{Present address: D\'{e}partement de Physique et Regroupement Qu\'{e}b\'{e}cois sur les Mat\'{e}riaux de Pointe, Universit\'{e} de Sherbrooke, Sherbrooke, Qu\'{e}bec, Canada, J1K2R1}
\author{M.~H.~Devoret}
\affiliation{Departments of Applied Physics and Physics, Yale University, New Haven, CT 06520}
\author{S.~M.~Girvin}
\affiliation{Departments of Applied Physics and Physics, Yale University, New Haven, CT 06520}
\author{R.~J.~Schoelkopf}
\affiliation{Departments of Applied Physics and Physics, Yale University, New Haven, CT 06520}

\begin{abstract}
Superconducting circuits are promising candidates for constructing quantum bits (qubits) in a quantum computer; single-qubit operations are now routine\cite{Devoret:2004r,You:2005k}, and several examples\cite{Yamamoto:2003t,Berkley:2003r,Majer:2005u,Steffen:2006q,Hime:2006f,Ploeg:2007r,Niskanen:2007l} of two qubit interactions and gates having been demonstrated.
These experiments show that two nearby qubits can be readily coupled with local interactions. Performing gates between an arbitrary pair of distant qubits is highly desirable for any quantum computer architecture, but has not yet been demonstrated. An efficient way to achieve this goal is to couple the qubits to a quantum bus, which distributes quantum information among the qubits. Here we show the implementation of such a quantum bus, using microwave photons confined in a transmission line cavity, to couple two superconducting qubits on opposite sides of a chip. The interaction is mediated by the exchange of virtual rather than real photons, avoiding cavity induced loss. Using fast control of 
the qubits to switch the coupling effectively on and off, we demonstrate coherent transfer of quantum states between the qubits. The cavity is also used to perform multiplexed control and measurement of the qubit states. This approach can be expanded to more than two qubits, and is an attractive architecture for quantum information processing on a chip. 
\end{abstract}

\maketitle

There are several physical systems in which one could realize a quantum bus. A particular example is trapped ions\cite{Cirac:1995f,Leibfried:2003k} in which a variety of quantum operations and algorithms have been performed using the quantized motion of the ions (phonons) as the bus. Photons are another natural candidate as a carrier of quantum information\cite{Duan:2001y,Chou:2007r}, because they are highly coherent and can mediate interactions between distant objects.
To create a photon bus, it is helpful to utilize the increased interaction strength provided by the techniques of cavity quantum electrodynamics, where an atom is coupled to a single cavity mode.
In the strong coupling limit\cite{Mabuchi:2002q} the interaction is coherent, permitting the transfer of quantum information between the atom and the photon. 
Entanglement between atoms has been demonstrated with Rydberg atom cavity QED\cite{Hagley:1997q,Zheng:2000r,Osnaghi:2001l}.
Circuit QED\cite{Blais:2004r} is a realization of the physics of cavity QED with superconducting qubits coupled to a microwave cavity on a chip. Previous circuit QED experiments with single qubits have achieved\cite{Wallraff:2004r} the strong coupling limit and have demonstrated\cite{Houck:2007l} the transfer of quantum information from qubit to photon.
Here we perform a circuit QED experiment with two qubits strongly coupled to a cavity, and demonstrate a coherent, non-local coupling between the qubits via this bus.

Operations with multiple superconducting qubits have been performed and are a subject of current research. The first solid-state quantum gate has been demonstrated with charge qubits\cite{Yamamoto:2003t}.
For flux qubits, two-qubit coupling\cite{Majer:2005u} and a controllable coupling mechanism have been realized\cite{Hime:2006f,Ploeg:2007r,Niskanen:2007l}. Two phase qubits have also been successfully coupled\cite{Berkley:2003r} and the entanglement between them has been observed\cite{Steffen:2006q}. 
All of these interactions have been realized by connecting qubits via lumped circuit elements, with capacitive coupling in the case of charge and phase qubits, and inductive coupling for flux qubits.
Therefore, these coupling mechanisms have been restricted to local interactions and couple only nearest neighbor qubits. In this work, we present a coupling that is realized with a cavity, which is a distributed circuit element, rather than with the lumped elements used previously. The interaction between the qubits occurs via photons in the cavity; hence, the cavity acts as an interaction bus allowing a non-local coupling of multiple qubits.

\begin{figure}
\includegraphics{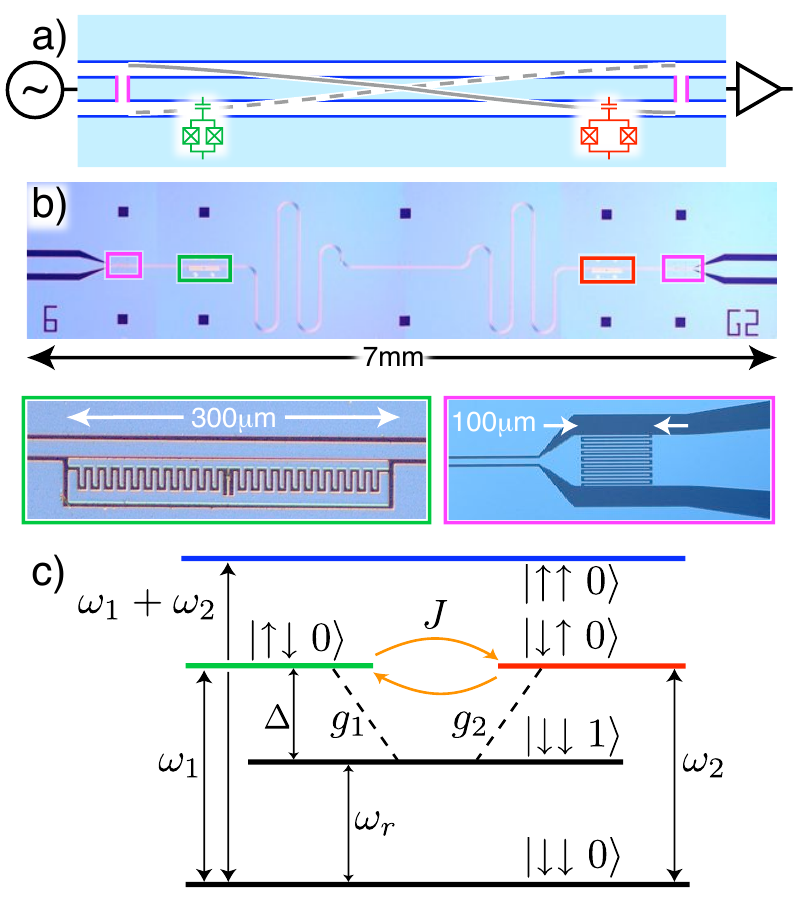}
\caption{
Sample and scheme used to couple two qubits to an on-chip microwave cavity.
Circuit \textbf{a} and optical micrograph \textbf{b} of the chip with two transmon qubits coupled by a microwave cavity.
The cavity is formed by a coplanar waveguide (light blue) interrupted by two coupling capacitors (purple). The resonant frequency of the cavity is $\omega_r/2\pi=5.19\:\mathrm{GHz}$ and its width is $\kappa/2\pi=33\:\mathrm{MHz}$, determined be the coupling capacitors. The cavity is operated as a half-wave resonator ($L=\lambda/2=12.3\:\mathrm{mm}$) and the electric field in the cavity is indicated by the gray line. 
The two transmon qubits (optimized Cooper-pair boxes) are located at opposite ends of the cavity where the electric field has an antinode. 
Each transmon qubit consists of two superconducting islands connected by a pair of Josephson junctions and an extra shunting capacitor (interdigitated finger structure in the green inset). The left qubit (qubit 1) has a charging energy of ${E_C}_1/h=424\:\mathrm{MHz}$ and maximum Josephson energy of ${E_J}_1^\mathrm{max}/h=14.9\:\mathrm{GHz}$. The right qubit (qubit 2) has a charging energy of ${E_C}_2/h= 442\:\mathrm{MHz}$ and maximum Josephson energy of ${E_J}_2^\mathrm{max}/h=18.9\:\mathrm{GHz}$.
The loop area between the Josephson junctions for the two transmon qubits differs by a factor of approximately $5/8$, allowing a differential flux bias.
The microwave signals enter the chip from the left, and the response of the cavity is amplified and measured on the right.
\textbf{c} Scheme of the dispersive qubit-qubit coupling. When the qubits are detuned from the cavity ($\left |\Delta_{1,2}  \right | =\left |\omega_{1,2} -\omega_r\right | \gg  g_{1,2}$) the qubits both dispersively shift the cavity. The excited state in the left qubit $\left|\uparrow\downarrow 0\right>$ interacts with the excited state in the right qubit $\left|\downarrow\uparrow 0\right>$ via the exchange of  a virtual photon $\left|\downarrow\downarrow 1\right>$ in the cavity.
\label{SchemePicture}
}
\end{figure}

To realize the cavity bus, we place two superconducting qubits 5 mm apart at opposite ends of a superconducting transmission line resonator (Fig.\ \ref{SchemePicture}a, \ref{SchemePicture}b).
The qubits are transmons\cite{Koch:2007l}, a modified version of the Cooper-pair box. In this type of qubit, the Josephson energy is larger than the charging energy ($E_J\gg E_C$) and the transition frequency between the ground state and the first excited state is given by $\omega\approx\sqrt{8 E_J E_C}/\hbar$. The Josephson junctions are arranged in a split-pair geometry, so that the Josephson energy, $E_J=E_J^\mathrm{max}\left|\cos(\pi\Phi/\Phi_0)\right|$ depends on the magnetic flux $\Phi$ applied through the split-pair loop. Hence, the transition frequency of the qubits, $\omega_{1,2}=\omega_{1,2}^\mathrm{max}\sqrt{\left|\cos(\pi\Phi/\Phi_0)\right|}$ can be tuned \textit{in-situ} with the applied flux. The size of the two loops is different and incommensurate, so that control of the two transition frequencies is attainable with a certain degree of independence. 
To probe the state of the system, homodyne detection of the transmitted signal is performed and both quadrature voltages are recorded, which allows reconstruction of the phase and amplitude of the transmitted signal.

\begin{figure}
\includegraphics{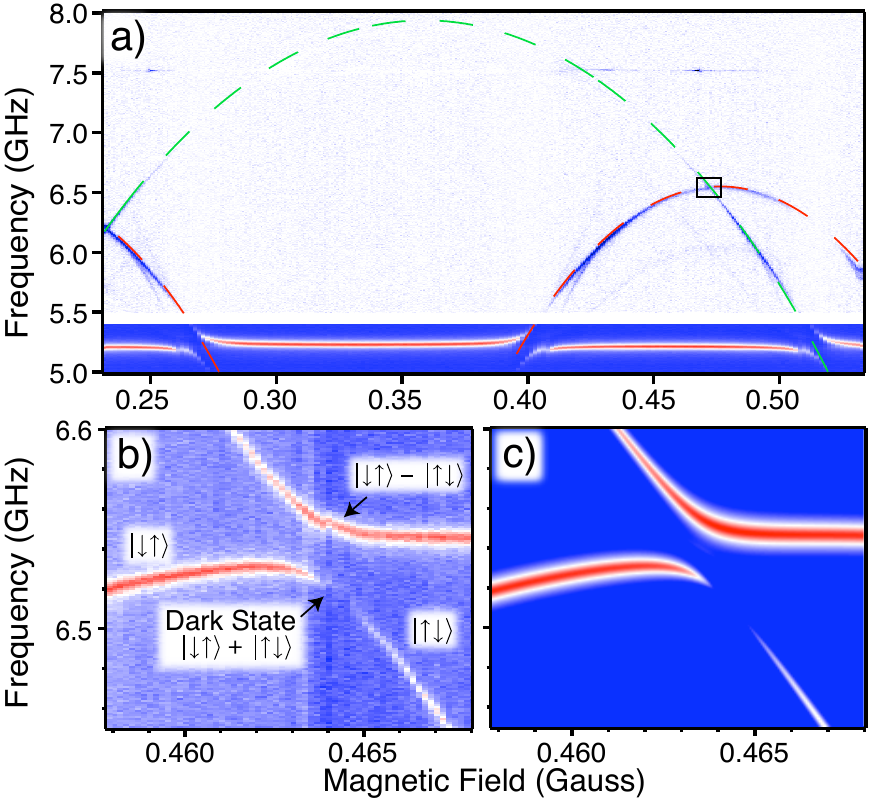}
\caption{Cavity transmission and spectroscopy of single and coupled qubits.
\textbf{a} The transmission through the cavity as a function of applied magnetic field is shown in the frequency range between 5~GHz and 5.4~GHz. When either of the qubits is in resonance with the cavity, the cavity transmission shows an avoided crossing due to the vacuum Rabi splitting. 
The maximal vacuum Rabi splitting for the two qubits is the same within the measurement uncertainty and is $\sim$105~MHz.
Above 5.5~GHz, spectroscopic measurements of the two qubit transitions are displayed. A second microwave signal is used to excite the qubit and the dispersive shift of the cavity frequency is measured.
The dashed lines show the resonance frequencies of the two qubits, which are a function of the applied flux according to $\omega_{1,2}=\omega_{1,2}^\mathrm{max}\sqrt{\left|\cos(\pi\Phi/\Phi_0)\right|}$.
The maximum transition frequency for the first qubit is $\omega_{1}^\mathrm{max}/2\pi=7.8\:\mathrm{GHz}$  and for the second qubit is $\omega_{2}^\mathrm{max}/2\pi=6.45\:\mathrm{GHz}$. 
For strong drive powers, additional resonances between higher qubit levels are visible.
\textbf{b} Spectroscopy of the two-qubit crossing. The qubit levels show a clear avoided crossing with a minimal distance of $2J/2\pi=26\:\mathrm{MHz}$. At the crossing the eigenstates of the system are symmetric and anti-symmetric superpositions of the two qubit states. The spectroscopic drive is anti-symmetric and therefore unable to drive any transitions to the symmetric state, resulting in a dark state. 
\textbf{c} Predicted spectroscopy at the qubit-qubit crossing using a Markovian master equation that takes into account higher modes of the cavity. The parameters for this calculation are obtained from the vacuum Rabi splitting and the single qubit spectroscopy. 
\label{TransmissionSpec}
}
\end{figure}

\begin{figure*}
\includegraphics{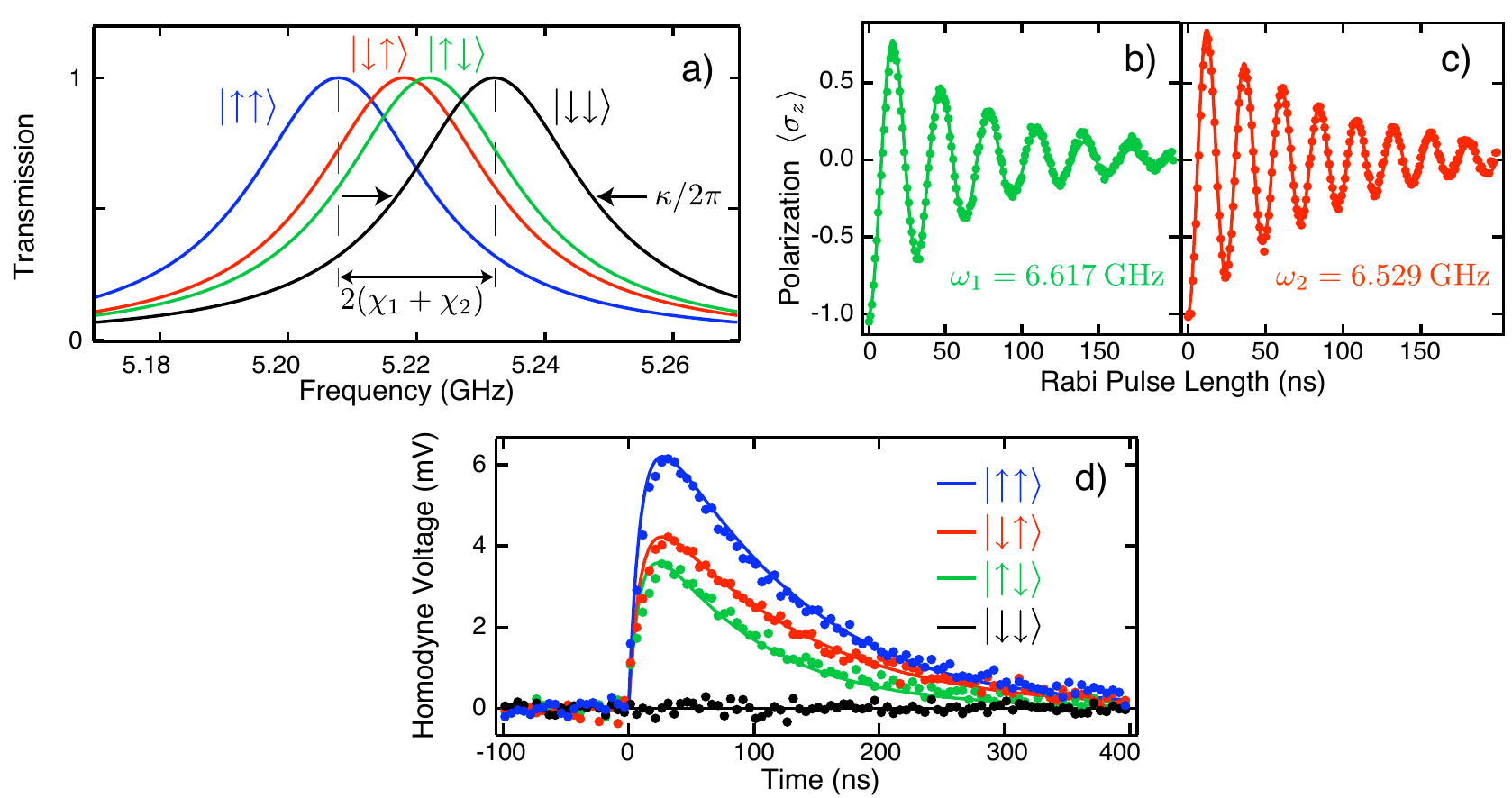}
\caption{Multiplexed control and read-out of uncoupled qubits.
\textbf{a} Predicted cavity transmission for the four uncoupled qubit states. 
In the dispersive limit ($\left |\Delta_{1,2}  \right | =\left |\omega_{1,2} -\omega_r\right | \gg  g_{1,2}$), the frequency is shifted by $\chi_{1}\sigma_1^z + \chi_{2}\sigma_2^z$. Operating the qubits at transition frequencies $\omega_1/2\pi = 6.617\:\mathrm{GHz}$ and $\omega_2/2\pi = 6.529\:\mathrm{GHz}$, we find $\chi_1/2\pi=-5.9\:\mathrm{MHz}$ and $\chi_2/2\pi=-7.4\:\mathrm{MHz}$. Measurement is achieved by placing a probe at a frequency where the four cavity transmissions are distinguishable. The two-qubit state can then be reconstructed from the homodyne measurement of the cavity.
Rabi oscillations of \textbf{b} qubit 1 and \textbf{c} qubit 2. A drive pulse of increasing duration is applied at the qubit transition frequency and the response of the cavity transmission is measured after the pulse is turned off.  Oscillations of quadrature voltages are measured for each of the qubits and mapped onto the polarization $\langle \sigma^{z}_{1,2} \rangle$. The solid line shows results from a master equation simulation, which takes into account the full dynamics of the two qubits and the cavity.
The absence of beating in both traces is a signature of the suppression of the qubit-qubit coupling at this detuning. 
\textbf{d} The homodyne response (average of 1,000,000 traces) of the cavity after a $\pi$ pulse on qubit 1 (green), qubit 2 (red), and both qubits (blue). The black trace shows the level when no pulses are applied. 
The contrasts\cite{Wallraff:2005v}(i.e.\ the amplitude of the pulse relative to its ideal maximum value) for these pulses are 60\% (green), 61\% (green) and 65\% (blue). The solid line shows the simulated value including the qubit relaxation and the turn-on time of the cavity. The agreement between the theoretical prediction and the data indicates the measured contrast is the maximum observable. 
From the theoretical calculation one can estimate the selectivity (see text for details) for each $\pi$-pulse to be 87\% (qubit 1) and 94\% (qubit 2). We note that this figure of merit is not at all intrinsic and that it could be improved by increasing the detuning between the two qubits for instance, or using shaped excitation pulses.
\label{MultiplexedControl}
}
\end{figure*}

In the first measurement we observe strong coupling of each of the qubits separately to the cavity.
By varying the flux, each of the two qubits can be tuned into resonance with the cavity (see Fig.\ \ref{TransmissionSpec}a). Whenever a qubit and the cavity are degenerate, the transmission is split into two well-resolved peaks in frequency, an effect called vacuum Rabi splitting\cite{Wallraff:2004r}, demonstrating that each qubit is in the strong coupling limit with the cavity. Each of the peaks corresponds to a superposition of qubit excitation and a cavity photon in which the energy is shared between the two systems.
From the frequency difference at the maximal splitting, the coupling parameters $g_{1,2}\approx 105\:\mathrm{MHz}$ can be determined for each qubit.
The transition frequency of each of the two qubits (see Fig.\ \ref{TransmissionSpec}a) can also be measured far from the cavity frequency as described below.

In the remainder of the experiment we operate the system in the dispersive limit,
where both qubits are detuned from the resonator $\left( |\Delta_{1,2}|=|\omega_{1,2}-\omega_r|\gg g_{1,2}\right)$.
In this limit, we use second order perturbation theory and the full system with the two qubits and the cavity is described by the effective Hamiltonian\cite{Blais:2007d}:
\begin{eqnarray*}
H_\mathrm{eff}&=&\frac{\hbar \omega_1}{2}\sigma^z_1+\frac{\hbar \omega_2}{2}\sigma^z_2+\hbar\left(\omega_r +\chi_1\sigma^z_1+\chi_2\sigma^z_2\right)a^\dag a\\
&&+\hbar J\left(\sigma_1^-\sigma_2^++\sigma_2^-\sigma_1^+\right)
\label{Hamiltonian}
\end{eqnarray*}
In this regime, no energy is exchanged with the cavity. However, the qubits and cavity are still dispersively coupled, resulting in a qubit-state-dependent shift $\pm\chi_{1,2}$ of the cavity frequency (see Fig.\ \ref{MultiplexedControl}a) or equivalently an AC Stark shift of the qubit frequencies\cite{Schuster:2005e}. The frequency shift $\chi_{1,2}$ can be calculated from the detuning $\Delta_{1,2}$ and the measured coupling strength $g_{1,2}$\cite{Koch:2007l}. The last term describes the interaction between the qubits, which is a transverse exchange interaction of strength $J=g_1 g_2(1/\Delta_1+1/\Delta_2)/2$ (See Fig.\ \ref{SchemePicture}c). The qubit-qubit interaction is a result of virtual exchange of photons with the cavity. When the qubits are degenerate with each other, an excitation in one qubit can be transferred to the other qubit by virtually becoming a photon in the cavity (see Fig.\ \ref{MultiplexedControl}b). However, when the qubits are non-degenerate $|\omega_1-\omega_2|\gg J$ this process does not conserve energy, and therefore the interaction is effectively turned off. Thus, instead of modifying the actual coupling constant\cite{Hime:2006f,Ploeg:2007r,Niskanen:2007l}, we control the \textit{effective} coupling strength by tuning the qubit transition frequencies. This is possible since the qubit-qubit coupling is transverse, which also distinguishes our experiment from the situation in liquid-state NMR quantum computation, where an effective switching-off can only be achieved by repeatedly applying decoupling pulses\cite{Gershenfeld:1997r}.

\begin{figure*}
\includegraphics{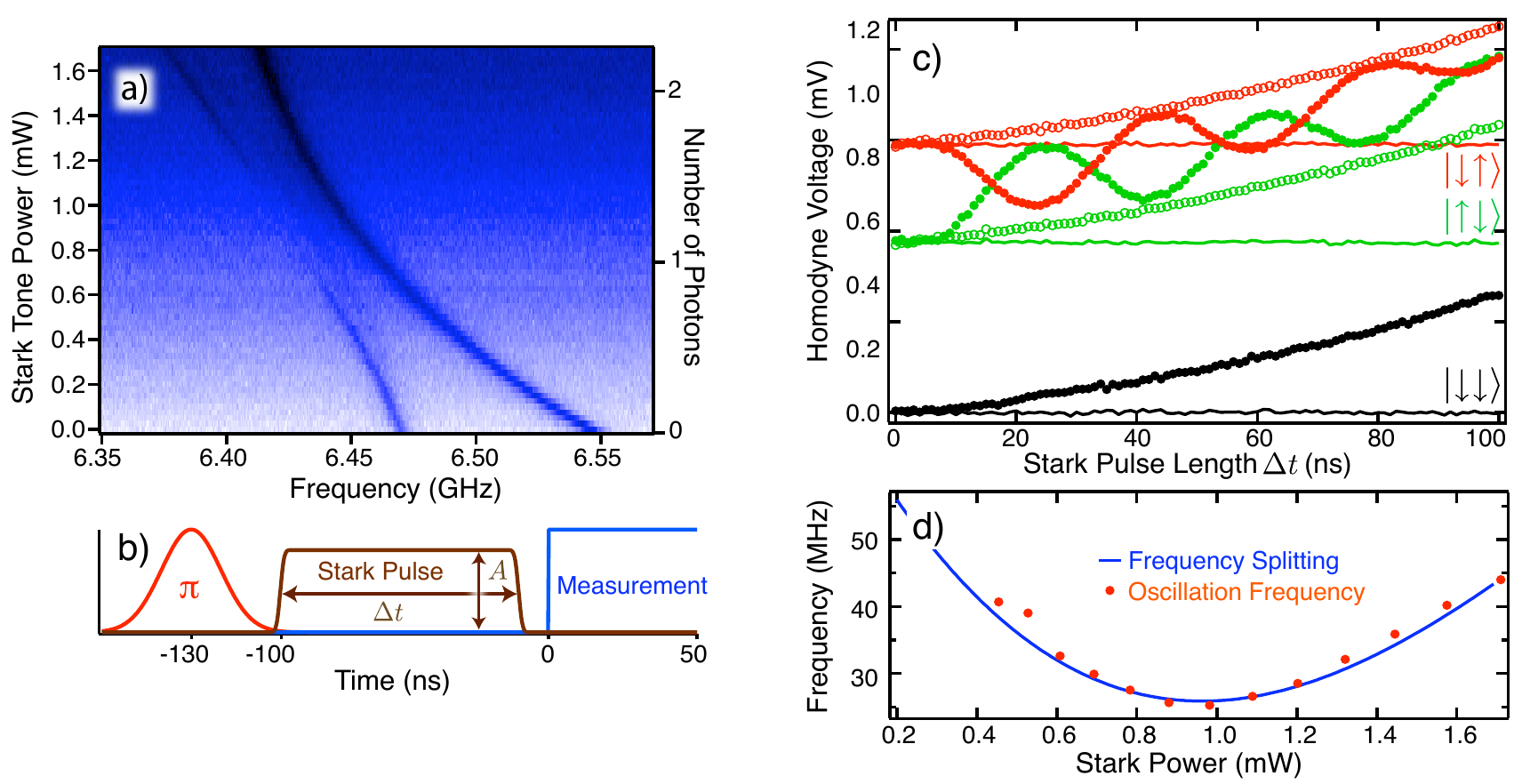}
\caption{
Controllable effective coupling and coherent state transfer via off-resonant Stark shift.
\textbf{a} Spectroscopy of qubits versus applied Stark tone power. Taking into account an attenuation of 67 dB before the cavity and the filtering effect of the cavity, 0.77 mW corresponds to an average of one photon in the resonator. The qubit transition frequencies (starting at $\omega_1/2\pi = 6.469\:\mathrm{GHz}$ and  $\omega_2/2\pi =  6.546\:\mathrm{GHz}$) are brought into resonance with a Stark pulse applied at $6.675\:\mathrm{GHz}$. An avoided crossing is observed with one of the qubit transition levels becoming dark as in Figure \ref{TransmissionSpec}.
\textbf{b} Protocol for the coherent state transfer using the Stark shift. The pulse sequence consists of a Gaussian-shaped $\pi$ pulse (red) on one of the qubits at its transition frequency $\omega_{1,2}$ followed by a Stark pulse (brown) of varying duration $\Delta t$ and amplitude $A$ detuned from the qubits, and finally a square measurement pulse (blue) at the cavity frequency. The time between the $\pi$ pulse and the measurement is kept fixed at 130~ns.
\textbf{c} Coherent state transfer between the qubits according to the protocol above. The plot shows the measured homodyne voltage (average of 3,000,000 traces) with the $\pi$ pulse applied to qubit 1 (green dots) and to qubit 2 (red dots) as a function of the Stark pulse  length $\Delta t$. For reference, the black dots show the signal without any $\pi$ pulse applied to either qubit. The overall increase of the signal is caused by the residual Rabi driving due to the off-resonant Stark tone, which is also reproduced by the theory. Improved designs featuring different coupling strengths for the individual qubits could easily avoid this effect. The thin solid lines show the signal in the absence of a Stark pulse. Adding the background trace (black dots) to these, we construct the curves consisting of open circles, which correctly reproduce the upper and lower limits of the oscillating signals due to coherent state transfer.
\textbf{d} The oscillation frequency (red) of the time domain state transfer measurement (c) and the splitting frequency (blue) of the continuous wave spectroscopy (a) versus power of the Stark tone. The agreement shows that the oscillations are indeed due to the coupling between the qubits.
\label{StarkSwap}
}
\end{figure*}

We first observe the coherent interaction between the two qubits via the cavity by performing spectroscopy of their transition frequencies (see Fig.\ \ref{TransmissionSpec}). 
This is done by monitoring the change in cavity transmission when the qubits are probed by a second microwave signal. By applying a magnetic flux the qubits can be tuned through resonance with each other (see Fig.\ \ref{TransmissionSpec}b), revealing an avoided crossing.
The magnitude of the splitting agrees well with the theoretical value $2J=2g_1g_2/\Delta= 2\pi\cdot 26\:\mathrm{MHz}$ when one takes into account that $g_{1,2}$ vary with frequency for a transmon qubit\cite{Koch:2007l}. The splitting is well resolved, with a magnitude $J$ much greater than the qubit line widths, indicating a coherent coupling and that the qubits are in the strong dispersive limit\cite{Schuster:2007h}. Note that although the coupling strength $J$ is smaller than the cavity decay rate $\kappa/2\pi\sim33\:\mathrm{MHz}$, the avoided crossing is nearly unaffected by the cavity loss. This is possible in such a large-$\kappa$ cavity, required for fast measurements, because only virtual photons are exchanged; if real photons were used, the cavity induced relaxation of the qubits (Purcell effect\cite{Houck:2007l}) would make coherent state transfer unfeasible.

Another manifestation of the coherence of this interaction is the observation of a dark state.  
One observes a disappearance of the spectroscopic signal near the crossing point, which is due to destructive interference associated with the fact that the qubits are separated by half a wavelength. At the crossing, the eigenstates are superpositions of the single qubit states. In particular, the state with lower frequency is the symmetric triplet state $\left|\downarrow\uparrow\right>+\left|\uparrow\downarrow\right>$ and the state at higher frequency is the antisymmetric singlet state $\left|\downarrow\uparrow\right>-\left|\uparrow\downarrow\right>$. In the dispersive limit, the spectroscopic excitation is of the form $\sigma^x_1\;g_1/\Delta_1-\sigma^x_2\;g_2/\Delta_2$, where the negative sign is due to the opposite signs of the electric field at the different ends of the $\lambda/2$ cavity as shown in Fig.\ \ref{SchemePicture}a, \ref{SchemePicture}b. 
Thus, such an external signal applied to the cavity cannot drive any transitions to the symmetric state, and is therefore dark.
Moreover, just as the triplet state does not couple to the drive, it is protected against decay through the cavity. 
Conversely, the decay from the singlet state is enhanced, similar to super-radiant effects observed in atomic physics\cite{Grangier:1985f,Itano:1998r}.
Figure \ref{TransmissionSpec}c shows the calculated spectroscopy at the qubit-qubit crossing, which  reproduces all qualitative features of the measured data.

In addition to acting as a quantum bus, the cavity can also be used for multiplexed read-out and control of the two qubits. Here, ``multiplexed" refers to acquisition of information or control of more than one qubit via a single channel.
To address the qubits independently, the flux is tuned such that the qubit frequencies are  88~MHz apart ($\omega_1= 6.617\:\mathrm{GHz}, \omega_2=6.529\:\mathrm{GHz}$), making the qubit-qubit coupling negligible.
Rabi experiments showing individual control are performed by applying an rf-pulse at the resonant frequency of either qubit, followed by a measurement pulse at the resonator frequency. 
The response (see Fig.\ \ref{MultiplexedControl}b and \ref{MultiplexedControl}c) is consistent with that of a single qubit oscillation and shows no beating, indicating that the coupling does not affect single-qubit operations and read-out.  
With similar measurements the relaxation times ($T_1$) of the two qubits are determined to be 78~ns and 120~ns, and with Ramsey measurements the coherence times ($T_2$) are found to be 120~ns and 160~ns.
The ability to simultaneously read-out the states of both qubits using a single line is shown by measuring the cavity phase shift, proportional to $\chi_1\sigma^z_1+\chi_2\sigma^z_2$ (see Eq.\ \ref{Hamiltonian}), after applying a $\pi$-pulse to one or both of the qubits.
Figure \ref{MultiplexedControl}d shows the response of the cavity after a $\pi$-pulse has been applied on the first qubit (green points), on the second qubit (red points) or on both qubits (blue points). For comparison the response of the cavity without any pulse applied (black points) is shown. Since the cavity frequency shifts for the two qubits are different ($\chi_1\neq\chi_2$), so we are able to distinguish the four states $\left|\downarrow\downarrow\right>,\left|\downarrow\uparrow\right>,\left|\uparrow\downarrow\right>,\left|\uparrow\uparrow\right>$ of the qubits with a single read-out line. 
One can show that this measurement, with sufficient signal to noise and combined with single-qubit rotations, should in principle allow for a full reconstruction of the density matrix (state tomography)\cite{Steffen:2006q}, although not demonstrated in the present experiment.

The solid lines in Figure \ref{MultiplexedControl}d show the results from a theoretical calculation taking into account the full dynamics of the cavity and the two qubits, including the relaxation in the qubits. The agreement of the theory with the measured response shows that the measured contrast is the maximum expected. 
From the calculated values one can estimate the selectivity, i.e.\ the ability to address one qubit without affecting the other, $S=(P_a-P_u)/(P_a+P_u)$, where $P_a$ and $P_u$ are the maximum populations in the excited state of the addressed qubit and in the excited state of the unaddressed qubit, respectively. The selectivity for qubit 1 is 87~\% and qubit 2 is 94~\%, which indicates good individual control of the qubits.

We can perform coherent state transfer in the time domain by rapidly turning the effective qubit-qubit coupling on and off.
Rather than the slow flux tuning discussed above, we now make use of a strongly detuned rf-drive\cite{Blais:2007d}, which results in an off-resonant Stark shift of the qubit frequencies on the nanosecond time scale. 
Figure \ref{StarkSwap}a shows the spectroscopy of the two qubits when this off-resonant Stark drive is applied with increasing power. The qubit frequencies are pushed into resonance and a similar avoided crossing is observed as in Fig.\ \ref{TransmissionSpec}b.
With the Stark drive's ability to quickly tune the qubits into resonance, it is possible to observe coherent oscillations between the qubits, using the following protocol (see Fig.\ \ref{StarkSwap}b):
Initially the qubits are 80~MHz detuned from each other, where their effective coupling is small, and they are allowed to relax to the ground state  $\left|\downarrow\downarrow\right>$. 
Next, a $\pi$-pulse is applied to one of the qubits to either create the state $\left|\uparrow\downarrow\right>$ or $\left|\downarrow\uparrow\right>$. Then a Stark pulse of power $P_\mathrm{AC}$ is applied bringing the qubits into resonance for a variable time $\Delta t$. Since $\left|\uparrow\downarrow\right>$ and $\left|\downarrow\uparrow\right>$ are not eigenstates of the coupled system, oscillations between these two states occur, as shown in Fig.\ \ref{StarkSwap}c. Fig.\ \ref{StarkSwap}d shows the frequency of these oscillations for different powers $P_{AC}$ of the Stark pulse, which agrees with the frequency domain measurement of the frequency splitting observed in Fig.\ \ref{StarkSwap}a. These data are strong evidence that the oscillations are due to the coupling between the qubits and that the state of the qubits is transferred from one to the other.
A quarter period of these oscillations should correspond to a $\sqrt{i\mathrm{SWAP}}$, which would be a universal gate. Future experiments will seek to demonstrate the performance and accuracy of this state transfer.

The observed qubit-qubit avoided crossing and the coherent state transfer demonstrate that the cavity can act as a coupling bus for superconducting qubits. The interaction is coherent and effectively switchable; furthermore, the coupling is long range, can easily be extended to non-nearest neighbors,  and it is protected against loss in the bus by the use of virtual photons. This architecture is not restricted to two qubits as there is room to couple many more qubits to the cavity, opening up new possibilities for quantum information processing on a chip.

\begin{acknowledgments}
This work was supported in part by the National Security Agency under the Army Research Office, by the NSF, and by Yale University. J.\ M.\ C.\ acknowledges support from an NSF Graduate Research Fellowship. A.\ A.\ H.\ acknowledges support from Yale University via a Quantum Information and Mesoscopic Physics Fellowship. L.~F.\ acknowledges partial support from CNR-Istituto di Cibernetica, Pozzuoli, Italy.
\end{acknowledgments}

\pagebreak

Correspondence and requests for materials should be addressed to Johannes Majer (email: {jo\nolinebreak[4]hannes.majer@yale.edu}) and  Robert Schoelkopf (email: robert.schoelkopf@yale.edu).

\bibliography{cQED_Coupling_condmat_final.bib}

\end{document}